\documentclass[pra,twocolumn]{revtex4-1}
\usepackage{graphicx,color}% Include figure files
\usepackage{epsfig}
\usepackage{graphicx}
\usepackage{mathrsfs}
\usepackage{color}
\usepackage{bm}
\usepackage{amsmath,amssymb,amsthm,amscd}
\usepackage{mdwlist}

\newcommand{\Tr}{\operatorname{Tr}}
\begin{document}

\title{Convex approximations of quantum channels}
\author{Massimiliano F. Sacchi}
\affiliation{Istituto di Fotonica e Nanotecnologie - CNR, Piazza Leonardo
  da Vinci 32, I-20133, Milano, Italy}
\affiliation{QUIT Group, Dipartimento di Fisica, 
Universit\`a di Pavia, via A. Bassi 6, I-27100 Pavia, Italy}
\author{Tito Sacchi}
\affiliation{Istituto Comprensivo di via Acerbi, via E. Acerbi 21, I-27100 Pavia, Italy}
\date{\today}

\begin{abstract}
We address the problem of optimally approximating the action of a desired and 
unavailable quantum channel $\Phi $ having at our disposal a single use of 
a given set of other channels 
$\{\Psi _i \}$. The problem is recast to look for the least distinguishable channel from $\Phi $
among the convex set $\sum _i p_i \Psi _i$, and the corresponding optimal weights $\{ p_i \}$ 
provide the optimal convex mixing of the available channels $\{\Psi _i \}$.  
For single-qubit channels we study specifically the cases  
where the available convex set corresponds to covariant channels or to Pauli channels, 
 and the desired target map is 
an arbitrary unitary transformation or a generalized damping channel. 
\end{abstract}

\maketitle 
\section{Introduction}
Quantum channels, or trace-preserving completely positive maps,
represent all possible deterministic quantum operations one can
perform over a quantum system \cite{NC}.  It is well known that
dilation theorems prove that a quantum channel can be realized by
means of a unitary transformation which couples the system to an
ancilla with fixed state preparation \cite{Kraus,busc}.  This
``realization'' theorem is clearly a powerful theoretical instrument,
but in a realistic scenario where the available technology is limited
it may give just a poor indication about the best effective
experimental realization of a desired quantum channel.

In the same spirit adopted to characterize the
universality of fixed quantum gates for quantum computation \cite{NC}, 
to study programmable devices to achieve different quantum channels
\cite{prog1,prog2,prog22,prog222} or measurements \cite{prog3,prog4}, and 
to perform the purification of noisy quantum measurements \cite{noise},   
in this paper we address the problem of optimally approximating the action of a
desired and unavailable quantum channel $\Phi $ over a Hilbert space
${\sf H}$ by an operational approach, when only a given set of quantum
channels $\{ \Psi _i\}$ for ${\sf H }$ is at our disposal for a single use.  More
specifically, we want to look for the best convex combination among
the channels of the given set that mostly resembles the desired $\Phi
$, i.e. that is the least distinguishable from $\Phi $ itself.  This
approach clearly has an immediate experimental application, especially
when the quantum operations effectively feasible in a lab are limited
due to intrinsic restrictions, unavailable technology, or even economic reasons.  Further
relevance of this approach is due to the fact that a convex sum of
quantum channels offers the possibility of performing  different
experiments followed by post-processing of experimental data
\cite{mataloni,mataloni2}, when the quantities of interest are linear with respect to quantum 
operations.

We note that  when the target channel is
unital, and the available set corresponds to all possible unitary
transformations, our problem is related to the quantification of the
distance between unital maps and random-unitary channels \cite{birk} 
and to the disproved ``quantum Birkoff's conjecture''
\cite{birk2,birk3}.

\section{Optimal convex approximation of quantum channels} 
The probability $p_{\mbox{\scriptsize discr}}$  of optimally discriminating 
between two quantum channels $\Phi_0 $ and $\Phi _1$ 
is quantified by the expression \cite{discrcp}
\begin{eqnarray}
p_{\mbox{\scriptsize discr}}
(\Phi _0, \Phi _1) =\frac 12 +\frac 14 ||\Phi _0 -\Phi _1|| _\diamond
\;,\label{uno}
\end{eqnarray}
where $|| \cdot ||_\diamond $ denotes the completely bounded trace norm \cite{paulsen}
(or, equivalently, the diamond norm \cite{diam}).  

By defining the positive Choi operator \cite{choi}
$R_{\Phi }= (\Phi \otimes I)|\eta \rangle  \langle \eta |$, 
which corresponds to the action of the map $\Phi $ 
over one party of a maximally entangled vector $|\eta \rangle \equiv 
\sum_{n=1}^d |n\rangle \otimes  |n\rangle$ of ${\sf H}\otimes {\sf H}$, with $d=\mbox{dim} (\sf H)$,  
let us recall the following identity \cite{discrcp}
\begin{eqnarray}
&&\Vert \Phi _0 -\Phi _1 \Vert  _\diamond = 
\nonumber \\& & 
\max _ {\mbox{\scriptsize $\Tr [\xi^\dag \xi ] =1$}} 
\left 
\Vert (I\otimes \xi ) (R_{\Phi _0}- R_{\Phi _1}) (I \otimes \xi ^\dag ) \right \Vert _1
\;,\label{pest}
\end{eqnarray}
where $\Vert  A \Vert _1 $ denotes the trace norm of $A$, namely \cite{bhatia} 
\begin{eqnarray}
\Vert A \Vert _1= \hbox{Tr}\sqrt{A^\dag A}= \sum _i s_i (A)\;, 
\end{eqnarray}
$\{s_i (A)\}$ representing the singular values of $A$. In the case of
Eq. (\ref{pest}), since the operator inside the norm is Hermitian,
the singular values just correspond to the absolute value of the
eigenvalues.  We also notice that any operator $\xi $ providing the maximum in Eq. (\ref{pest}) 
corresponds to 
an optimal input state $(I\otimes \xi )|\eta \rangle $ for the discrimination \cite{nota}. 
We recall here also the result in Refs. \cite{acin,dlp}, namely for arbitrary unitary maps 
${\cal V} \equiv V  (\cdot )V^\dag $ and 
${\cal Z} \equiv Z  (\cdot ) Z^\dag $ one has 
\begin{eqnarray}
\Vert {\cal V}  - {\cal Z} \Vert  _\diamond = 2 \sqrt {1- r(Z^\dag V)^2} 
\;, 
\end{eqnarray}
where $r(Z^\dag V)$ 
denotes the distance from the origin of the complex plane of the polygon whose vertices 
stay on the circle of unit radius and correspond to 
the singular values of the unitary matrix $Z^\dag V$.  
For $d=2$, since there are 
only two eigenvalues, one easily finds by simple geometry that  
$r(U) =\frac 12 |\Tr [U]|$.

The problem of the optimal convex approximation of a quantum channel is implicitly posed by 
the following definition. 

{\bf Definition:} The optimal convex approximation of a quantum channel $\Phi $ 
w.r.t. a given set of quantum channels $\{ \Psi _i \}$ is given by 
$\sum _ i p_i ^{opt}\Psi _i$,  where $\{ p_i^{opt}\}$ denotes the vector of probabilities
\begin{eqnarray}
\{p_i ^{opt}\}= \arg \min _ {\{ p_i\}} 
\Vert \Phi - \textstyle\sum _i p_i \Psi _i  \Vert _\diamond 
\;.  
\end{eqnarray}
The effectiveness of the optimal convex approximation 
is then quantified by the $\{ \Psi _i\}$-{\em distance}
\begin{eqnarray}
D_{\{\Psi _i \}}(\Phi )\equiv \min _ {\{ p_i\}} \Vert 
\Phi - \textstyle\sum _i p_i \Psi _i \Vert _\diamond 
\;,  
\end{eqnarray}
which provides through Eq. (\ref{uno}) the worst probability 
of discriminating the desired channel $\Phi $ from any of the available channels 
$\sum _ i p_i \Psi _i$. Clearly, our definition of optimal convex
approximation can be suitably changed by referring to any other figure
of merit that quantifies the distance between quantum operations \cite{fid1,fid2}. 

Since we have in mind an operational approach where the channels in the set 
$\{\Psi _i \}$ are experimentally available, 
we always assume that this  set contains the identity map $\cal I$. We note that 
the formulation of the diamond norm as a semidefinite program satisfying strong duality 
\cite{spd,spd2,spd3} 
allows its efficient calculation. 
Moreover, the convexity of the norm itself allows one  
to search for the minimum by means of standard software of convex optimization \cite{cvx,cvx2}. 

From the convexity of the diamond norm, it follows the upper bound 
\begin{eqnarray}
D_{\{\Psi _i \}}(\Phi )\leq \min _ {i} ||\Phi - \Psi _i ||_\diamond = \min _i 
D_{ \Psi _i }(\Phi )
\;.\label{upp}
\end{eqnarray}
On the other hand, since from Eq. (\ref{pest}) one has  
\begin{eqnarray}
\Vert \Phi _0 -\Phi _1 \Vert  _\diamond \geq 
\frac 1d 
\left 
\Vert R_{\Phi _0}- R_{\Phi _1} \right \Vert _1
\;,\label{pest2}
\end{eqnarray}
one obtains the lower bound 
\begin{eqnarray}
D_{\{\Psi _i \}}(\Phi )\geq \frac 1d \min _ {\{ p_i\}} 
\left 
\Vert R_{\Phi } - R_{\sum _i p_i \Phi _i} \right \Vert _1
\;. 
\end{eqnarray}
From the unitarily invariance of the diamond norm, notice also that for all  
unitary maps  ${\cal V} $
and ${\cal Z}$ 
one has the symmetry
\begin{eqnarray}
D_{\{ \Psi _i \}}(\Phi )=D_{ \{ {\cal V}\circ \Psi _i \circ {\cal Z} \} }
({\cal V}\circ \Phi \circ {\cal Z})
\;, 
\end{eqnarray}
where $\circ $ denotes the composition of maps. 
Clearly, if the set itself $\{ \Psi _i\}$ is invariant, then  
\begin{eqnarray}
D_{\{\Psi _i \}}({\cal V}\circ \Phi \circ {\cal Z})=
 D_{\{ \Psi _i \}}(\Phi )
\;,\label{inva}
\end{eqnarray}
and the probabilities of the optimal convex approximation for  
${\cal V}\circ \Phi \circ {\cal Z}$ are just a permutation of those for $\Phi $. 
This is the case, for example, when the available channels are unitary maps corresponding to 
a (projective) representation of some elements of a group. 

\section{Distance of a unitary map from covariant channels} 
Let us consider the case where the set of available channels is given by ${\cal C}
=\{ {\cal I}, \frac{1}{d^2-1}(d\Tr[\cdot ] I -{\cal I}) \}$. The convex hull 
is clearly given by the channels 
\begin{eqnarray}
{\cal C}_p (\rho )= (1-p) \rho + \frac{p}{d^2-1}(d \Tr[\rho] I -\rho ) 
\;,\label{covp}
\end{eqnarray}
with $p\in [0,1]$. Indeed, Eq. (\ref{covp}) describes all covariant channels 
for $SU(d)$, namely the channels $\cal E $ satisfying 
\begin{eqnarray}
U^\dag _g {\cal E}[U_g \rho U_g ^\dag ]U_g ={\cal E}(\rho)\;\label{covv}
\end{eqnarray}
for all $\rho $ and unitary $U_g \in SU(d)$ \cite{notaff}. 
Notice also that Eq. (\ref{covp}) includes all depolarizing channels 
(for $p\in [0,(d^2-1)/d^2] $). 
For any orthogonal basis of unitaries $\{V_i\}$ containing the identity $V_0\equiv I$ 
(and hence with  
$\Tr[V_i]=0$ and $\Tr [V^\dag _i V_ j]=d \delta _{ij}$ for $i,j=1,\ldots,d^2-1$) we have also 
${\cal C}= \{ {\cal I}, \frac{1}{d^2-1}\sum _{i=1} ^{d^2-1} V_i (\cdot ) V_i ^\dag  \}$. 
This means that the optimal convex approximation of a channel $\Phi $ with respect to 
covariant channels can be achieved by the convex mixture of the identity map and equally-weighted 
orthogonal rotations.  
\par Let us now study the case of qubits, where the target map $\Phi $ 
is a unitary transformation, which, up to a global phase,  can be parameterized as 
\begin{eqnarray}
U(\alpha ,\beta ,\gamma )=\left( 
\begin{array}{cc}
\cos \alpha e^{i\beta } & \sin \alpha  e^{i\delta } \\ 
-\sin \alpha e^{-i \delta }&
\cos\alpha e^{-i \beta } \\
\end{array}
\right )
\;,\label{uabd}
\end{eqnarray}
with $\alpha \in [0,\pi/2]$ and $\beta\in [0,2\pi]$, and $\delta \in [0,2\pi]$. 
Denote, as usual, the Pauli matrices as $\sigma _0=I$, $\sigma _1=\sigma _x$, 
$\sigma _2 =\sigma _y$, and 
$\sigma _3=\sigma _z$. The convex hull of the set $\{{\cal I} , 
\frac 13 \sum_{i=1}^3 \sigma_i  (\cdot )\sigma _i \}$ 
provides all covariant channels as in Eq. (\ref{covp}), for $d=2$.  
Of course, the convex approximation of the map 
${\cal U}(\alpha ,\beta, \delta ) =U(\alpha ,\beta ,\gamma   )(\cdot )
U^\dag (\alpha ,\beta ,\gamma   )$ 
is rather poor, since the set is highly constrained. 
However, we can give here a complete analytical solution, and 
the physical 
interpretation of the result is crystalline and exemplary for more intricate situations. 

Then, for qubits, 
the {\em covariance distance} of the unitary map ${\cal U}(\alpha ,\beta , \delta ) $
is given by  
\begin{eqnarray}
D_{\cal C}[{\cal U}(\alpha ,\beta , \delta ) ]
\equiv \min _{p\in [0,1]} \Vert {\cal U}(\alpha ,\beta , \delta ) 
- {\cal C}_p \Vert _\diamond \;.\label{udep}
\end{eqnarray}
Since the difference of the Choi operators $R_{\cal U}$ and $R_{{\cal C}_p}$ 
can be diagonalized over orthonormal Bell states \cite{discrcp}, one obtains 
\begin{eqnarray}
&&\Vert {\cal U}(\alpha ,\beta , \delta ) - {\cal C}_p 
\Vert _\diamond = 
\frac 12 \left
\Vert R_{\cal U}- R_{{\cal C}_p} \right \Vert _1  \label{udepno}
\\& & =
\frac 23 p + \sqrt{\frac{16}{9} p^2 + 
\left (1-\frac 43  p\right ) D_I[{\cal U}(\alpha ,\beta , \delta )  ]^2}
\;,\nonumber 
\end{eqnarray}
where $D_I[{\cal U}(\alpha ,\beta ,\delta )]=
\Vert {\cal U}(\alpha ,\beta ,\delta ) -{\cal I} \Vert _\diamond =
2(1- \cos^2 \alpha \cos^2 \beta )^{1/2}$.  
Equation (\ref{udepno}) provides via Eq. (\ref{uno}) 
 the minimum-error discrimination between arbitrary 
unitary channels and covariant channels.  
No maximization is present in Eq. (\ref{udepno}) 
because any maximally entangled state can always be used as an input to achieve the optimal discrimination, 
thus allowing one to choose $\xi =\frac {I} {\sqrt 2}$ in Eq. (\ref{pest}).   
\par The minimization over $p$ of Eq. (\ref{udepno}) 
gives for the covariance distance the piecewise function
\begin{eqnarray}
&& D_{\cal C}[{\cal U}(\alpha ,\beta ,\delta ) ]=  \label{piece}
\\& & 
 \left\{
\begin{array}{ll}
x
&   \quad 0\leq x 
\leq 1 \\
\frac 14 [ x^2 + \sqrt { 3 x^2 (4-x^2)}] & \quad 1\leq x\leq 
\left ( \frac{15+\sqrt {33}}{6}\right )^{1/2} \\
\frac 13 (2+ \sqrt{16-3 x^2}) & \quad \left ( \frac{15+\sqrt {33}}{6} \right )^\frac 12 
\leq x \leq 2
\end{array}
\right. \;
\nonumber 
\end{eqnarray}
where, for brevity, we posed $x\equiv D_I[{\cal U}(\alpha ,\beta ,\delta )]$. 
The minimum is achieved for $p=0$, $p=\frac{1}{8}[3 x^2 -\sqrt{3 x^2(4 - x^2)}]$, 
and $p=1$ in the three respective pieces. Being just a function of 
$D_I [{\cal U}(\alpha ,\beta ,\delta )]$,  
the covariance distance 
$D_{\cal C}[{\cal U}(\alpha ,\beta ,\delta ) ]$ is independent of the parameter $\delta $. 

In Fig. 1 the covariance distance $D_{\cal C}[{\cal U}(\alpha ,\beta,\delta )]$ 
for unitary maps is plotted vs. the diamond norm $D_I
[{\cal U}(\alpha ,\beta ,\delta )]$. 
It is quite easy to physically interpret the result: 
as long as the unitary $U(\alpha ,\beta ,\delta )$ 
is close enough to the identity (i.e. $x\leq 1$) 
nothing can be done to approximate its action, whereas, 
for $U(\alpha ,\beta ,\delta )$ sufficiently far from the identity,  
the optimal convex approximation is an equally-weighted rotation by the three Pauli matrices. 
In fact, in these two cases the upper 
bound in Eq. (\ref{upp}) is saturated with equality. 
Finally, in between these two situations, one has to suitably weight the two previous strategies
 with probability  $p=\frac{1}{8}[3 x^2 -\sqrt{3 x^2(4 - x^2)}]$.

\begin{figure}[htb]
  \includegraphics[width=\columnwidth]{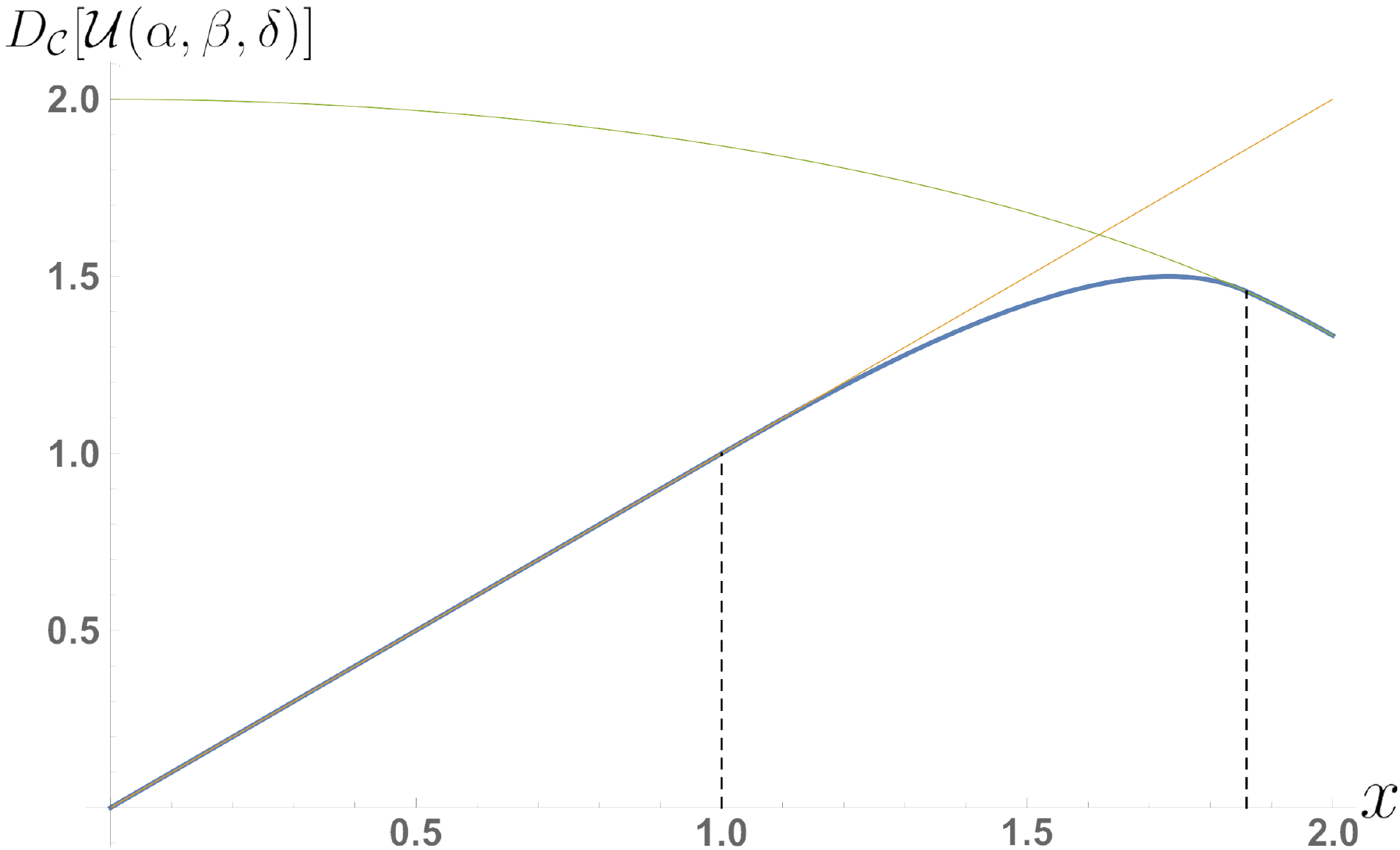}
  \caption{Optimal convex approximation of a unitary map ${\cal
  U}(\alpha, \beta ,\delta )$ for qubits w.r.t. the set including the
  identity map and the map $\frac 13 \sum_i \sigma _i  (\cdot )\sigma _i $.  The problem is
  solved by finding the closest covariant channel to the unitary map in
  the diamond norm, as in Eq. (\ref{udep}). The covariance distance
  $D_{\cal C}[{\cal U}(\alpha ,\beta,\delta )]$ is
  a piecewise function of the diamond norm $x\equiv D_{I}[{\cal U}(\alpha
  ,\beta,\delta )]= \Vert {\cal U}(\alpha ,\beta ,\delta ) -{\cal I}
  \Vert _\diamond $, as given in Eq. (\ref{piece}).}  
  \label{fig:depo}
\end{figure}

\section{Pauli distance of a unitary map} 
For a set given by the identity map ${\cal V}_0 \equiv {\cal I}$ 
and $(d^2-1)$ unitary maps ${\cal V}_i$, 
with corresponding traceless and orthogonal unitary operators, 
the optimal convex approximation 
of a quantum channel $\Phi $ corresponds to the closest generalized Pauli channel, 
namely to a channel of the form $\sum _{i=0}^{d^2- 1} {p_i} V_i\rho V_i^\dag $ 
which provides the minimum {\em Pauli distance}
\begin{eqnarray}
D_{\cal P}(\Phi )\equiv D_{\{ {\cal V}_i\}} (\Phi )\;. 
\end{eqnarray}
Let us study in more detail the case of qubit channels, and consider $\Phi $ as the unitary map 
${\cal U}(\alpha ,\beta ,\gamma )$, using again the parametrization of Eq. (\ref{uabd}).  
Exploiting the invariance properties of the 
set of Pauli matrices, according to Eq. (\ref{inva}), 
a number of symmetry relations for the Pauli distance can be derived, which can be summarized 
as follows  \cite{ntt}
\begin{eqnarray}
&&D_{\cal P }[{\cal U}(\alpha ,\beta, \delta )] =
D_{\cal P}
\left [{\cal U} \left (\alpha , \frac \pi 2 \pm \beta , \delta  \right )\right ] 
= 
\label{sym}    \\& & 
D_{\cal P} \left [{\cal U}\left (\alpha ,\beta, \frac \pi 2 \pm \delta \right )\right ] =
D_{\cal P}
\left [{\cal U}\left (\frac \pi 2 - \alpha ,\delta , \beta \right )\right ] 
\;.\nonumber 
\end{eqnarray}
For specific unitaries $U(\alpha ,\beta ,\delta )$  
we can find exact results for the optimal convex approximation:
 
$i)$ 
for $ \beta = \delta =0 $ one has 
\begin{eqnarray}
D_{\cal P}[{\cal U}(\alpha ,0,0)] =|\sin 2 \alpha |  
\;, 
\end{eqnarray}
with pertaining optimal weights given by 
$\{p _i ^{opt}\}=\{\cos^2 \alpha ,0,\sin^2 \alpha ,0\}$; 
 
$ii)$ for $ \alpha =0$ one has 
\begin{eqnarray}
D_{\cal P}[{\cal U}(0,\beta ,\delta )] =|\sin 2 \beta |
\;, 
\end{eqnarray}
with $\{p _i ^{opt}\}
=\{\cos^2 \beta ,0,0, \sin^2 \beta \}$; 

$iii)$ for $ \alpha =\pi/2$ one has 
\begin{eqnarray}
D_{\cal P}[{\cal U}(\pi/2 ,\beta ,\delta ) =|\sin 2 \delta |
\;, 
\end{eqnarray}
with $\{p _i ^{opt}\}=\{0,\sin^2 \delta ,\cos ^2 \delta ,0 \}$. 

\begin{figure}[thb]
  \includegraphics[width=\columnwidth]{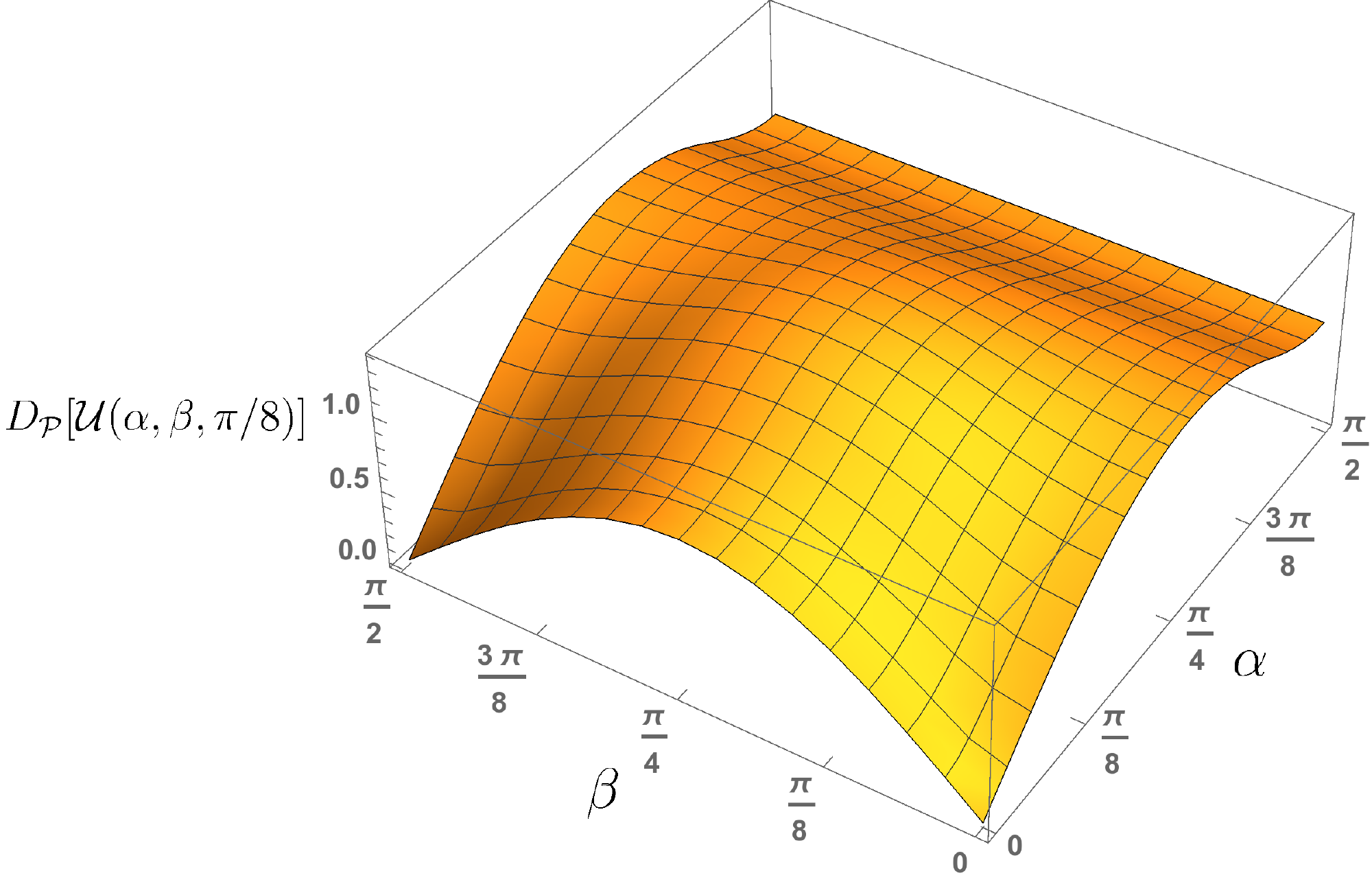}
  \caption{Optimal convex approximation of a unitary map ${\cal
  U}(\alpha, \beta ,\delta )$ w.r.t. the set including the
  identity map and the three rotations by the three Pauli matrices. The problem is
  solved by finding the closest Pauli channel to the unitary map in
  the diamond norm.
The solution provides the Pauli distance  
$D_{\cal P}[{\cal U}(\alpha ,\beta,\delta )]$, here plotted vs. $ \alpha  $ and $ \beta $, 
for a fixed value of $ \delta  $, namely $ \delta \equiv \pi /8$.} 
  \label{fig:depo2}
\end{figure}
Notice that in all the above specific examples the optimal vector of
probabilities has just two non-zero elements. More
generally, however, the optimal convex approximation requires three or
even all four $ \sigma _i $-operations.  For generic values of $ \alpha
, \beta, \delta $ one can look for a numerical solution. 
As an example, in Fig. 2 we present the result of the optimal convex approximation of 
the  unitary maps ${\cal U}(\alpha ,\beta ,\pi/8) $. 
The unitary maps which are worst approximated corresponds to 
$\alpha = \beta =\delta =\frac \pi 4 $ 
[along with those related by the symmetries in Eqs. (\ref{sym})], and their Pauli distance equals 
$\frac 32$ and is achieved for equal weights $p_i=\frac 14$, 
namely, by the completely depolarizing channel. 

\section{Pauli distance of a generalized damping channel} 
A generalized  damping channel $\Gamma (q,\gamma )$ for qubits is described by the  
completely positive map 
\begin{eqnarray}
\Gamma (q, \gamma )[\rho ] &=&q (A_\gamma \rho A _\gamma + C_\gamma \rho C_\gamma ^\dag) 
\nonumber \\& +& 
(1-q) (B_\gamma \rho B _\gamma + C_\gamma ^\dag  \rho C_\gamma  ) 
\;, 
\end{eqnarray}
where 
$A_\gamma =\left(%
\begin{array}{cc}
  1 & 0 \\
  0 & \sqrt{1-\gamma}\\
\end{array}%
\right) $, 
$B_\gamma =\left(%
\begin{array}{cc}
  \sqrt{1-\gamma} & 0 \\
  0 & 1 \\
\end{array}%
\right)$, and 
$C_\gamma=\left(%
\begin{array}{cc}
  0 & \sqrt{\gamma} \\
  0 & 0 \\
\end{array}%
\right)$, 
with $0\leq q,\gamma \leq 1$. 
This channel is a mixture of an amplitude damping channel ($q=1$) and an amplitude 
amplification channel ($q=0$), and thus $q$ plays the role of a temperature. 
\par Let us look for the optimal convex approximation of $\Gamma (q,\gamma )$ w.r.t. 
the set of Pauli matrices. 
From Eq. (\ref{inva}) we notice that the identity 
$\sigma _x \circ \Gamma (q,\gamma )\circ \sigma_x = \Gamma (1-q,\gamma )$  
provides the symmetry relation 
$D_{\cal P}[\Gamma  (q,\gamma )]=D_{\cal P}[\Gamma (1-q,\gamma )]$.  

\begin{figure}[hbt]
  \includegraphics[width=\columnwidth]{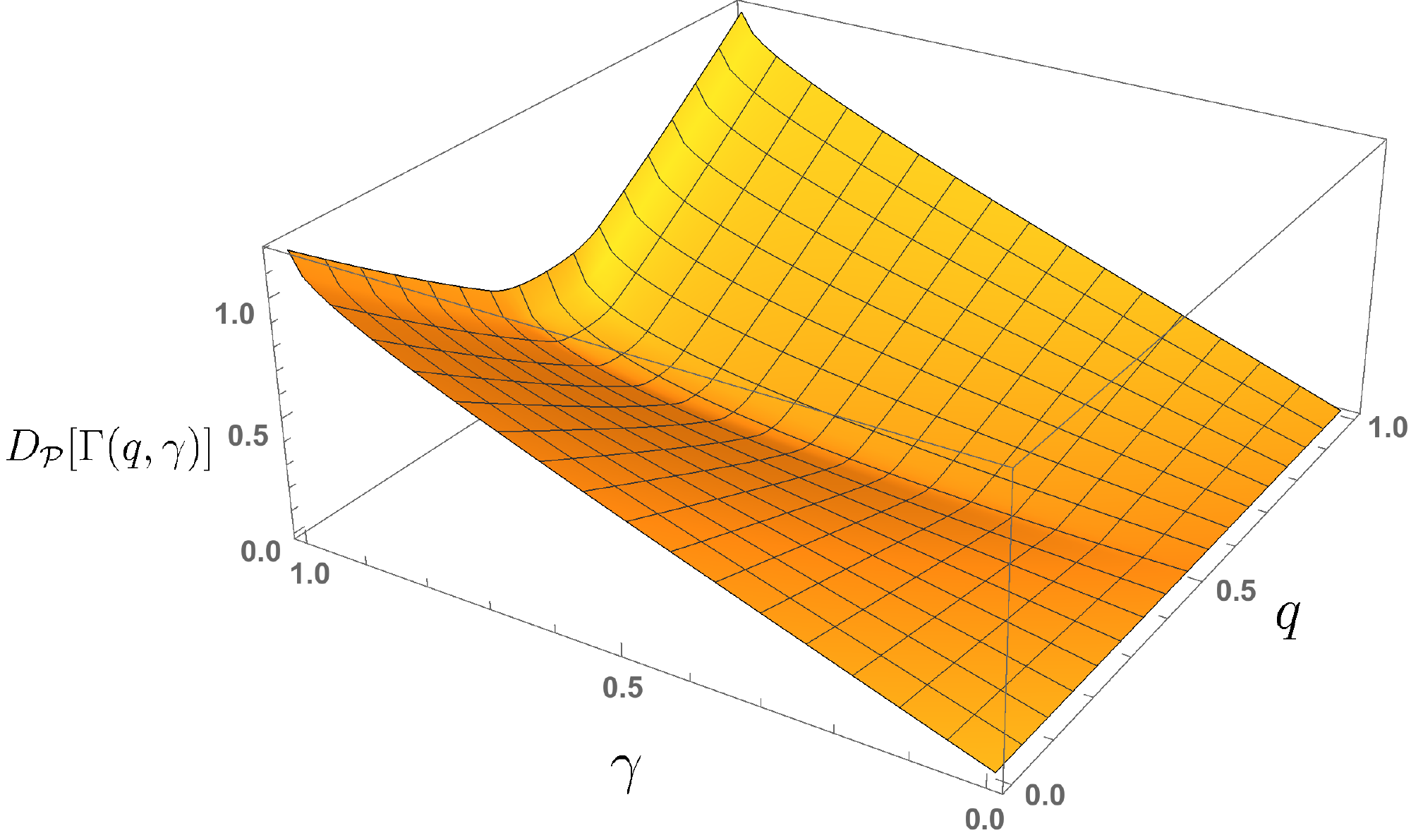}
  \caption{Optimal convex approximation of a generalized damping channel $\Gamma (q,\gamma)$ 
w.r.t. the set including the
  identity map and the three rotations by the three Pauli matrices. The problem is
  solved by finding the closest Pauli channel to $\Gamma (q,\gamma)$ in 
  the diamond norm. 
The solution provides the Pauli distance
  $D_{\cal P}[\Gamma (q,\gamma  )]
$, here plotted vs. $ q  $ and $ \gamma $.}
  \label{fig:damp}
\end{figure}

We have numerically solved the problem, and the results are plotted in Fig. 3, where  
we show the Pauli distance of $D_{\cal P}[\Gamma (q,\gamma )]$ vs. $q$ and $\gamma $. 
The pertaining weights $\{p_i^{opt}\}$ of the optimal Pauli approximation 
are of the form $\{1-2p,p,p,0\}$. 
On one hand, one has $p_1^{opt}=p_2^{opt}$, 
namely the rotations by $ \sigma _x $ and $\sigma _y $ are equally weighted. 
In fact, this condition guarantees that the 
convex approximation enjoys the same covariance property of $\Gamma (q, \gamma )$, i.e. 
\begin{eqnarray}
{\cal V}(\phi ) \circ \Gamma (q,\gamma ) =
\Gamma (q,\gamma ) \circ {\cal V}(\phi ) \qquad \forall \phi
\;, 
\end{eqnarray}
where ${\cal V}(\phi )=e^{i\phi \sigma _z}
(\cdot )e^{-i\phi \sigma _z}$ 
denotes the rotation map around $ \sigma _z $. 
On the other hand, the additional result $p_3^{opt}=0$ 
stems from the fact that any phase rotation by  
$\sigma _z $ would make the resulting Pauli channel more distinguishable from 
$\Gamma (q,\gamma)$, 
which instead preserves the phase of the off-diagonal matrix-elements of quantum states.  
We have numerical evidence that the optimal probability satisfies $p_1 ^{opt}
\geq \frac \gamma 4$, along with the following bounds for the Pauli distance 
\begin{eqnarray}
\!\!\!\gamma |1-2q|  
\leq  D_{\cal P}[\Gamma (q,\gamma  )] \leq  
\frac 12 [\gamma |1-2q| + f(q,\gamma)]
\,,\label{bb}
\end{eqnarray}
with the function $f(q,\gamma )$ given in Eq. (\ref{effe}) in the Appendix, 
where in fact these bounds are proved. 
\par\noindent In Fig. 4 we plot the Pauli distance for the channel $\Gamma [0.7, \gamma ]$ vs. 
the parameter $\gamma $, along with the upper and lower bounds. It is apparent that 
these bounds are tighter for decreasing values of $\gamma $, and the upper bound is indeed 
very good.   
\begin{figure}[tbh]
  \includegraphics[scale=0.65]{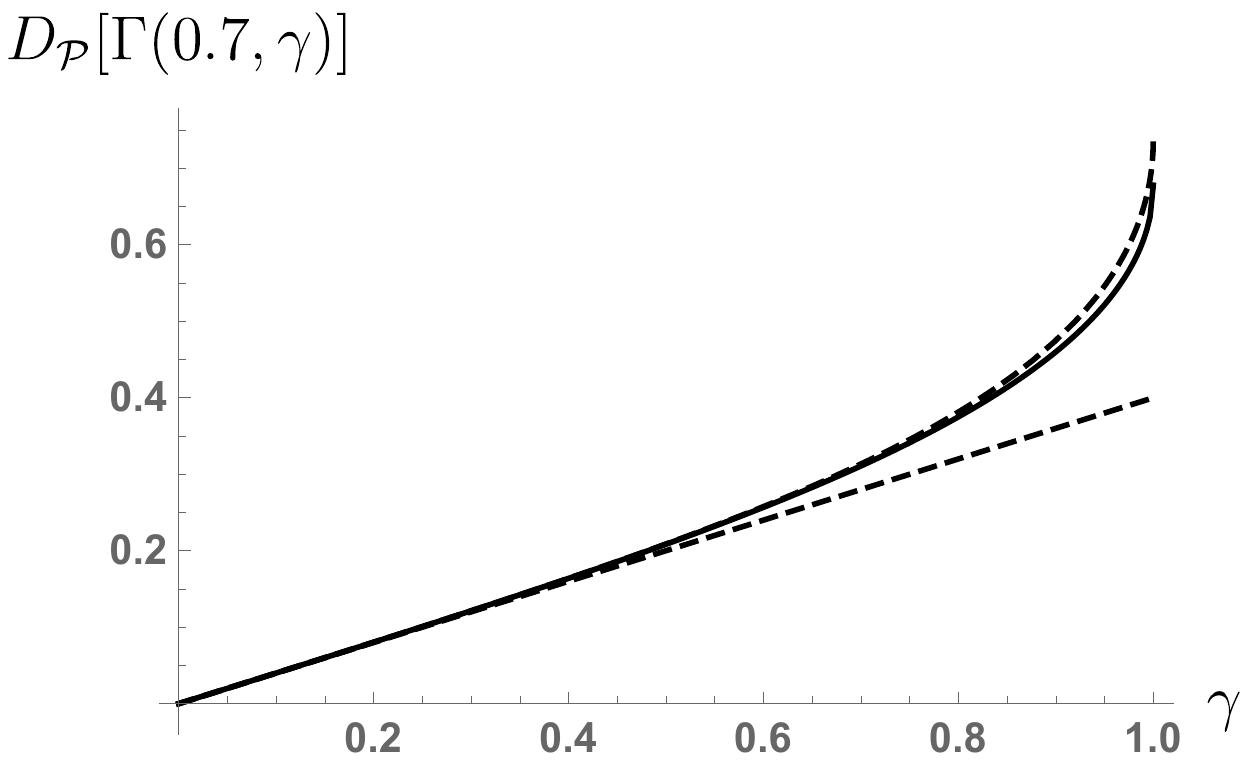}
  \caption{Optimal convex approximation of a generalized damping channel $\Gamma (0.7,\gamma)$ 
vs. the parameter $\gamma $,   
with respect to the set of Pauli channels. The Pauli distance
  $D_{\cal P}[\Gamma (0.7,\gamma  )]$ is plotted in solid line, along with 
the upper and lower bounds given by Eq. (25) in dashed line.}
\label{fig:damp2}
\end{figure}
 
\section{Conclusions} 
Let us conclude our paper with the following observations. 
Imagine that we want to 
approximate $N$ parallel uses of a map $\Phi $ acting on an (unknown) $N$-partite quantum state, 
and we have at our disposal a set of maps 
$\{\Psi _i\}$ with which we can act independently on each subsystem. 
The optimal convex approximation in this case provides the distance 
$D_{ \{ \otimes _{j=1}^N  \Psi _{i_j} \} }
(\Phi ^{\otimes N})$. 
Since obviously the convex hull of ${\{ \otimes _{j=1}^N  \Psi _{i_j} \}}$ contains all 
the $N$-fold tensor products $\otimes _{j=1}^N (\textstyle
\sum _i p_{i_j } \Psi _i )$, 
one has 
\begin{eqnarray}
D_{\{\otimes _{j=1}^N  \Psi _{i_j} \}} (\Phi ^{\otimes N}) 
&\leq & 
\min _{ \{ p_{i_j } \} } 
\Vert \Phi ^{\otimes N} - \otimes _{j=1}^N (\textstyle
\sum _i p_{i_j } \Psi _i ) \Vert _{\diamond }  
\nonumber \\& 
\leq & \Vert \Phi ^{\otimes N} -(\textstyle
\sum _i p^{opt }_i \Psi _i )^{\otimes N} \Vert _{\diamond }
\;,\label{ineq}
\end{eqnarray}
where $\{p^{opt}_i\}$ denotes the vector of probabilities pertaining to the optimal 
convex approximation of a single copy of the target map $\Phi $. 
The interesting fact is that generally both inequalities in Eq. (\ref{ineq}) can be strict, 
and a simple explicit example is provided in the following. 
\par\noindent  Consider the unitary map ${\cal U}$ corresponding to the phase rotation
\begin{eqnarray}
U=\left( 
\begin{array}{cc}
e^{i\frac \pi 6 } & 0 \\ 
0& e^{-i \frac \pi 6 } \\
\end{array}
\right )
\;,\label{u6}
\end{eqnarray}
and its convex approximation w.r.t. $\{ \cal I, \cal Z \}$, where ${\cal Z}= \sigma _z 
(\cdot )\sigma _z$. Thus, according to Eq. (21), 
the optimal convex approximation is the closest dephasing channel 
$\Psi (p) = p {\cal I} + (1-p) {\cal Z} $ to 
$\cal U$, and one has 
\begin{eqnarray}
D_{\{{\cal I}, {\cal Z} \}}({\cal U}) = \frac {\sqrt 3} 2\;,
\end{eqnarray}
with corresponding optimal weight $p^{opt}=\frac 34$. The diamond norm for the two-fold 
tensor product can be evaluated as
\begin{eqnarray}
\Vert {\cal U}^{\otimes 2} -\Psi (3/4) ^{\otimes 2} \Vert _\diamond \simeq 1.314 
\;.
\end{eqnarray}
A tiny improvement is found by looking for the closest tensor product of dephasing channels 
as  
\begin{eqnarray}
\min _{q_1,q_2 \in [0,1]}\Vert {\cal U}^{\otimes 2} - \Psi (q_1) \otimes \Psi (q_2) \Vert _
{\diamond}
\simeq 1.312\;,
\end{eqnarray}
with optimal weights $q_1^{opt}=q_2^{opt}\simeq 0.77$. 
By allowing correlations,  a larger improvement is obtained by 
the optimal convex approximation, and one has 
\begin{eqnarray}
&&D_{\{ {\cal I}\otimes {\cal I}, {\cal Z}\otimes {\cal I}, {\cal I}\otimes {\cal Z}, 
{\cal Z}\otimes {\cal Z} 
\}} ({\cal U} ^{\otimes 2}) =
\min _{\{p_{ij} \}}\Vert 
{\cal U}^{\otimes 2} 
\nonumber \\& & 
-
(p_{00} {\cal I}\otimes {\cal I} +p_{10} {\cal Z}\otimes {\cal I}
+p_{01} {\cal I}\otimes {\cal Z}+p_{11} {\cal Z}\otimes {\cal Z})
\Vert _\diamond 
\nonumber \\& & 
\simeq 1.281
\;, 
\end{eqnarray}
where the optimal weights are given by $p^{opt}_{00}\simeq 0.60 $, 
$p^{opt}_{10}=p^{opt}_{01}\simeq 0.20$, and $p^{opt}_{11}=0$. The resulting optimal convex 
approximation is obviously a channel with correlated uses. 

The first inequality in Eq. (\ref{ineq}) comes from the fact that the introduction of correlations in the 
approximating map can be beneficial even if the target map is indeed the product 
of independent maps (as happens, for example, in the optimal cloning of quantum states 
\cite{clon}).  
The second inequality is due to 
the fact that the distance $\Vert \Phi_0 -\Phi _1 \Vert _\diamond $ 
quantifying the distinguishability of two channels is not 
additive/multiplicative when considering multiple copies, namely, we clearly only know that 
$\Vert \Phi_0 -\Phi _1 \Vert _\diamond \leq 
\Vert \Phi_0^{\otimes N} -\Phi _1 ^{\otimes N}\Vert _\diamond $. 
This also implies that we do not have a direct expression for the scaling with $N$ of the 
distance between a quantum channel and its convex approximations. 
The results related to the quantum Chernoff bound for quantum states 
\cite{chern,chern2}, suitably generalized to the case of quantum channels, 
might be useful for a systematic study of the scaling of the optimal convex approximations 
with the number of uses. 

\appendix*
\section{Proof of the bounds in Eq. (25).}
Let us consider the use of input state $|0 \rangle   $ or  $|1 \rangle $ for discriminating a 
generalized amplitude damping channel $\Gamma (q,\gamma )$ from a Pauli channel ${\cal P}=
\sum _{i=0} ^3 p_i \sigma _i (\cdot ) \sigma _i $.  
Then the diamond norm is bounded as 
\begin{eqnarray}
\Vert \Gamma (q,\gamma ) -{\cal P}\Vert _\diamond \geq 
\Vert (\Gamma (q,\gamma ) -{\cal P}) |u \rangle \langle  u | \Vert _1  
\;, %\qquad u=0,1\;.
\end{eqnarray}
for $u=0,1$. A straightforward calculation gives
\begin{eqnarray}
&&
\!\!\!\!
\Vert  (\Gamma (q,\gamma ) -{\cal P}) |0 \rangle \langle  0 | \Vert _1 =
2 |\gamma (1- q) -(p_1+p_2)| 
\\& & 
\!\!\!\!\Vert (\Gamma (q,\gamma ) -{\cal P}) |1 \rangle \langle  1 | \Vert _1 =
2 |\gamma  q -(p_1+p_2)| 
\;.
\end{eqnarray}
Then  the Pauli distance $D_{\cal P}[\Gamma (q,\gamma )]$  of the damping channel 
can be bounded as 
\begin{equation}
\begin{split}
&D_{\cal P}[\Gamma (q,\gamma )] \equiv 
\min _{\{  p_i\}} \Vert \Gamma (q,\gamma ) -{\cal P} \Vert _\diamond  
\nonumber \\ & \geq  
\min _{\{ p_i\}}
\max \{  2 |\gamma (1- q) -(p_1+p_2)|, 2 |\gamma  q -(p_1+p_2)| \}  \nonumber \\&  = 
\gamma |1-2q|
\;, 
\end{split}
\end{equation}
where the minimum is achieved for $p_1=p_2=\frac{\gamma}4 $. This proves the lower bound in Eq. 
(25). 

The upper bound can be simply obtained by choosing the Pauli channel ${\cal P}_\gamma $ with 
$p_0=1- \frac {\gamma }{2}$, $p_1=p_2=\frac {\gamma }4$, and $p_3=0$. Then, obviously,  
\begin{eqnarray}
D_{\cal P}[\Gamma (q,\gamma )] \leq 
\Vert \Gamma (q,\gamma ) - {\cal P}_\gamma  \Vert _  \diamond  
\;. \label{dpd}
\end{eqnarray}
The diamond norm in Eq. (\ref{dpd}) can be explicitly evaluated as 
\begin{eqnarray}
\Vert \Gamma (q,\gamma ) - {\cal P}_\gamma  \Vert _  \diamond  = 
\frac 12 \left [\gamma |1-2q| + f(q,\gamma) \right ]
\;,
\end{eqnarray}
with 
\begin{eqnarray}
&&f(q,\gamma )=  \label{effe}\\ &&
\{8 (1-\gamma) -4 (2-q)\sqrt {1-\gamma} +\gamma ^2 [2-4q(1-q)] 
\}^{1/2}\;. \nonumber 
\end{eqnarray}


\begin{thebibliography}{99}
\bibitem{nota} From the unitarily invariance of the trace norm 
and the polar decomposition of $\xi $, the matrix $\xi $ can be taken as a 
positive matrix $P$. 
Moreover, for numerical maximization, positivity can be constrained by 
means of the Choleski decomposition, namely $P=T^\dag T$, with $T$ upper diagonal 
with positive diagonal entries. For qubits, a useful parametrization is also 
\begin{eqnarray}
P=\left( 
\begin{array}{cc}
\cos \theta_1 \cos \theta _2  & \frac {1}{\sqrt 2} \sin \theta _2  e^{i\phi } 
\\
\frac {1}{\sqrt 2} \sin \theta _2  e^{-i\phi } & \sin \theta_1 \cos \theta _2
\end{array}
\right )\;,  
\end{eqnarray}
with $\theta_1 \in [0,\pi/ 2]$, 
$\theta _2\in [0, \arctan \sqrt{(\sin 2 \theta_1)}]$, and 
$\phi \in [0, 2\pi]$. 
Finally, the formulation of the trace norm as a semidefinite program \cite{boyd} allows 
its efficient computation.
\bibitem{notaff} This can be easily proved since, in terms of the Choi operator 
${\cal R}_{\cal E}
$, Eq. (\ref{covv}) is equivalent to $[U_g \otimes U^* _g,{\cal R}_{\cal E}]=0$. 
The application of Shur's lemma for unitary irreducible representations forces the channel 
${\cal E}$ to the form as in Eq. (\ref{covp}).
\bibitem{ntt}The symmetry relations are easily obtained by the fact that the set of 
Pauli matrix is a closed algebra, along with the covariance property 
$\{ \sigma _i \}=\{ S \sigma _i S \}=\{ S \sigma _i S ^\dag \}$ up to global phases, 
with $S= 
\left( 
\begin{array}{cc}
1  & 0  
\\0
 & i \end{array}
\right )$.
\end{thebibliography}
\end{document}